\documentclass[twocolumn,showpacs,amsmath,amssymb,prb]{revtex4}

\usepackage{graphicx}% Include figure files
\usepackage{dcolumn}% Align table columns on decimal point
\usepackage{bm}% bold math

\begin{document}

\preprint{APS/123-QED}

\title{Magneto-transport properties of CeRu$_2$Al$_{10}$: Similarities to URu$_2$Si$_2$}
\author{$^{1}$Jiahao Zhang}
\author{$^{1}$Sile Hu}
\author{$^{1}$Hengcan Zhao}
\author{$^{1}$Pu Wang}
\author{$^{2}$A. M. Strydom}
\author{$^{1}$Jianlin Luo}
\author{$^{1,3}$Frank Steglich}
\author{$^{1}$Peijie Sun}
 \email{pjsun@iphy.ac.cn}
\affiliation{%
$^{1}$Beijing National Laboratory for Condensed Matter Physics, Institute of Physics, Chinese Academy of Sciences, Beijing 100190, China\\
$^{2}$Highly Correlated Matter Research Group, Physics Department, University of Johannesburg, P.O. Box 524, Auckland Park 2006, South Africa\\
$^{3}$Max Planck Institute for Chemical Physics of Solids, 01187 Dresden, Germany
}

\date{\today}

\begin{abstract}
We report on magneto-transport properties of the Kondo semiconducting compound CeRu$_2$Al$_{10}$, focusing on its exotic phase below $T_0$ = 27 K. In this phase, an excess thermal conductivity $\kappa$ emerges and is gradually suppressed by magnetic field, strikingly resembling those observed in the hidden-order phase of URu$_2$Si$_2$. Our analysis indicates that low-energy magnetic excitation is the most likely origin, as was also proposed for URu$_2$Si$_2$ recently, despite the largely reduced magnetic moments. Likewise, other transport properties such as resistivity, thermopower and Nernst effect exhibit distinct features characterizing the very different charge dynamics above and below $T_0$, sharing similarities to URu$_2$Si$_2$, too. Given the exotic nature of the ordered phases in both compounds, whether a unified interpretation to all these observations exists appears to be extremely interesting.
\end{abstract}

\pacs{Valid PACS appear here}% PACS, the Physics and Astronomy
\maketitle

CeRu$_2$Al$_{10}$ is one of the heavy-fermion (HF) materials showing timely interest due to a novel phase transition at an abnormally high temperature $T_{\rm 0}$ = 27 K emerging in a Kondo semiconducting phase \cite{strydom09,nishioka09}. An antiferromagnetic scenario, as originally proposed by one of the authors \cite{strydom09} based on magnetic and thermodynamic measurements, has been frequently argued to be incompatible with the large nearest neighbor distance between Ce ions, $\sim$ 5.2 \AA, and the de Gennes scaling of ordering temperatures of $R$Ru$_2$Al$_{10}$ ($R$ = rare earth). $\mu$SR and neutron powder diffraction experiments have revealed a long-ranged, collinear antiferromagnetic ordering of the Ce sublattice below $T_0$ with, however, a strongly reduced magnetic moment of 0.34$-$0.42 $\mu_{\rm B}$, and an unusual alignment along $c$ axis that is not the magnetic easy direction  \cite{Kha10, kato}. The microscopic origin of the magnetic ordering, which apparently goes beyond the conventional Ruderman-Kittel-Kasuya-Yosida (RKKY) interaction, remains elusive.
On the other hand, transport properties of CeRu$_2$Al$_{10}$ are intriguing as well. For instance, regardless of the formation of a partial gap over the majority of the Fermi surface with onset of the new phase, the electrical resistivity $\rho(T)$ ceases to be semiconducting-like and becomes metal-like upon cooling below $T_0$. Moreover, both thermopower $S(T)$ and thermal conductivity $\kappa(T)$ are accompanied by an extra peak at temperatures slightly below $T_0$ or deep inside the ordered phase \cite{strydom09}.

In this work, we investigate the magneto-transport properties of CeRu$_2$Al$_{10}$. Special attention is put on the excess thermal conductance emerging in the ordered phase below $T_0$ and its strong suppression upon applying magnetic field. Note that, strikingly similar behaviors have ever been observed for another heavy-fermion system, URu$_2$Si$_2$, in the hidden-order phase below $T_{\rm h}$ = 17.5 K \cite{behnia05,sharma}. Enhanced phononic thermal conductance arising from a freezing out of relevant scattering centers below $T_{\rm h}$ has been argued to be the origin \cite{behnia05,sharma}. This argument, however, is questionable in view of the strong field dependence and because of recent detailed experimental investigations on phonon dynamics by inelastic x-ray scattering measurements \cite{gardner}. The acoustic phonon modes do not change significantly upon cooling into the hidden-order phase, arguing against the phononic origin of the excess $\kappa$ below $T_{\rm h}$. We will show that our experimental results for CeRu$_2$Al$_{10}$, on the other hand, strongly support low-energy magnetic excitations to be at the center of these phenomena. Moreover, main features in the resistivity, thermopower and Nernst coefficient are able to be approached by taking into account two regimes of very different charge dynamics separated by $T_0$.  These, again, share similarities with those of URu$_2$Si$_2$.

Polycrystalline sample of CeRu$_2$Al$_{10}$ and its nonmagnetic homologue LaRu$_2$Al$_{10}$ were prepared by arc-melting the stoichiometric starting materials and following an annealing process in vacuum at 800$^{\rm o}$ for one week \cite{strydom09}. Power x-ray diffraction confirms the YbFe$_2$Al$_{10}$-type (orthorhombic, space group Cmcm, No. 63) crystal structure. The obtained lattice constants are $a$ = 9.1254 \AA, $b$ = 10.2791 \AA, and $c$ = 9.1876 \AA, which are in good agreement with the reported values. Electrical resistivity $\rho(T)$, thermal conductivity $\kappa(T)$, thermopower $S(T)$, and Nernst coefficient $\nu(T)$ were measured in the physical property measurement system (PPMS, Quantum Design) between 2 K and room temperature, using a sample with typical dimension 0.5$\times$2$\times$5 mm$^3$. Among these, the Nernst measurements were performed on a home-design sample puck with one chip resistor of 2000\,$\Omega$ as heater and one thin ($\phi$\,$=$\,25\,$\mu$m) chromel-AuFe$_{0.07\%}$ thermocouple for detecting the temperature gradient, as described in ref. \cite{cosb3}.

\begin{figure}
\includegraphics[width=0.92\linewidth]{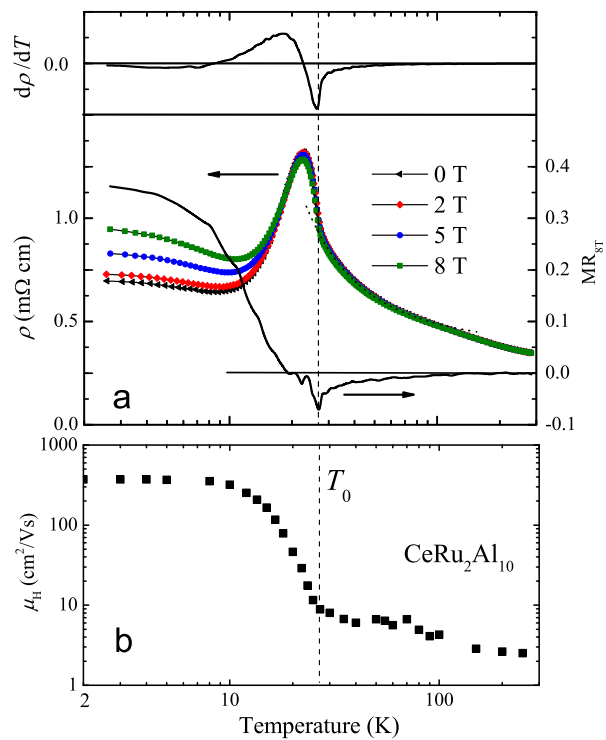}
\caption{(a) Electrical resistivity $\rho(T)$ measured in various magnetic fields applied perpendicular to electrical current, the magnetoresistance MR$_{\rm 8T}$ = ($\rho_{\rm 8T}$$-$$\rho_{\rm 0T}$)/$\rho_{\rm 0T}$, and the derivative of $\rho$  with respect to $T$ for $B$ = 0 T. Dotted line represents a thermal activation behavior of $\rho(T)$ observed between 30 and 80 K. (b) Hall mobility $\mu_{\rm H}$ as a function of temperature \cite{cosb3}.
\label{rho}}
\end{figure}

Figure~\ref{rho}a shows $\rho(T)$ measured in various magnetic fields applied perpendicular to electrical current, the derivative d$\rho$/d$T$, and the magnetoresistance MR($T$). In agreement with previous reports \cite{strydom09,nishioka09}, the curve of $\rho(T)$ displays a prominent maximum at $T$ $\approx$ 22 K well below $T_{\rm 0}$. The phase transition at $T_{\rm 0}$ is, however, clearly manifested by a sharp negative extreme in both d$\rho$/d$T$ and MR as a function of temperature. The value of MR($T$) is negative above $T_0$, evolving dramatically to be positive upon cooling. In order words, application of a magnetic field suppresses charge scattering events only at $T$ $>$ $T_0$. This points to the onset of a magnetically ordered phase at $T_0$, above which, spin fluctuations as scatterers of conduction electrons give rise to negative values of MR.

\begin{figure}
\includegraphics[width=0.92\linewidth]{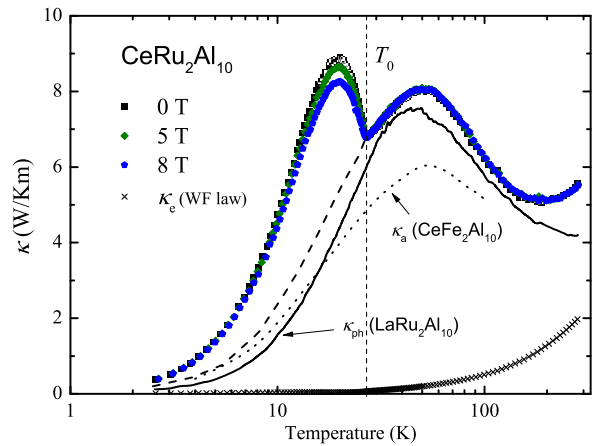}
\caption{Thermal conductivity $\kappa(T)$ measured in different magnetic fields for CeRu$_2$Al$_{10}$. The electronic contribution $\kappa_{\rm e}(T)$ (crosses) was estimated from the WF law on the basis of the measured values of resistivity in zero field. The solid line represents the phononic thermal conductivity $\kappa_{\rm ph}$ of the nonmagnetic analog LaRu$_2$Al$_{10}$, and the dotted curve $\kappa(T)$ of single-crystalline CeFe$_2$Al$_{10}$ along the most heat-conductive direction ($a$ axis). The dashed curve at temperatures below $T_0$ is an extrapolation of the measured $\kappa$ for CeRu$_2$Al$_{10}$, ignoring the excess contribution below $T_0$ and following essentially the same $T$ dependence of $\kappa_{\rm ph}$ of LaRu$_2$Al$_{10}$. By applying magnetic field, the excess heat conductance below $T_0$ is significantly reduced.
\label{kappa}}
\end{figure}

Given that a charge gap opens over nearly 90\% of the Fermi surface slightly above $T_0$ \cite{kimura,matsu09}, it is uncommon that $\rho(T)$ evolves from a semiconducting behavior above $T_0$ to be metal-like in the ordered phase. Interestingly, this unusual behavior shares some similarities to $\rho(T)$ of URu$_2$Si$_2$. Upon cooling down into the hidden-order phase, $\rho(T)$ of URu$_2$Si$_2$ passes through a maximum at a temperature below $T_{\rm h}$ and becomes more metallic at lower temperatures \cite{dawson}, even though the majority of the Fermi surface is gapped out. These features, observed in both compounds, apparently are related to the much larger Hall mobility $\mu_{\rm H}$ in the ordered phase relative to the paramagnetic phase (cf. Fig. 1b), and are consistent with the large, positive MR in the ordered phase. In a limited temperature range from above $T_0$ up to 80 K, $\rho(T)$ of CeRu$_2$Al$_{10}$ roughly follows a thermal activation law, with a small charge gap $E_g$ = 46 K that is comparable to that estimated from NMR \cite{matsu09} and optical spectra \cite{kimura}.

Figure~\ref{kappa} shows the measured $\kappa(T)$ for CeRu$_2$Al$_{10}$. It has two distinct maxima at $T$ = 20 and 50 K, separated by a sharp valley right at $T_0$. Such a temperature profile has been observed for both polycrystalline and single-crystalline samples \cite{strydom09, tanida10}, and the isoelectronic homologue CeOs$_2$Al$_{10}$ \cite{lue12} with a similar phase transition. A common practice to analyze thermal conductivity of a conducting solid assumes $\kappa$ to be the sum of a phononic and an electronic contribution, $\kappa = \kappa_{\rm e} + \kappa _{\rm ph}$. The electronic term $\kappa_{\rm e}$ can be readily estimated from the Wiedemann-Franz law, $\kappa_{\rm e}$$\rho$/$T$ = $L_0$, with the Sommerfeld value of Lorenz number $L_0$ $\equiv$ $\frac{\pi^2}{3}(\frac{k_B}{2})^2$ $=$ 2.44 $\times$ 10$^{-8}$ W$\cdot$$\Omega$$\cdot$K$^{-2}$. Due to the relatively large values of resistivity of CeRu$_2$Al$_{10}$,  $\kappa_{\rm e}$ is estimated to be negligibly small at $T$ $<$ 100 K, being easily ruled out as a relevant origin of the complex $\kappa(T)$ behavior.

\begin{figure}
\includegraphics[width=0.92\linewidth]{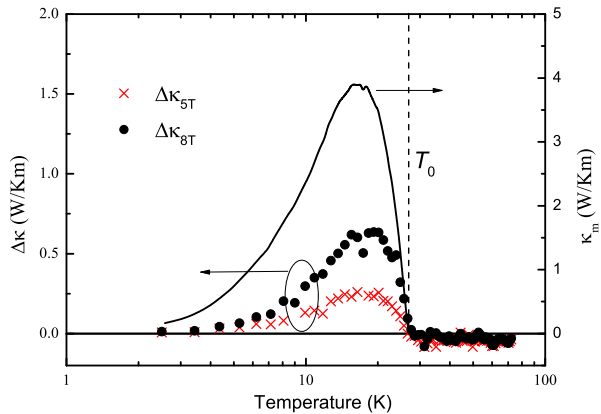}
\caption{The excess thermal conductivity $\kappa_{\rm m}$ in the ordered phase and its suppression in magnetic field, denoted as the difference of thermal conductivity measured in zero and applied magnetic fields, i.e.,  $\Delta \kappa$ = $\kappa_{\rm 0T}$ $-$ $\kappa_{\rm B}$. Note the appearance of non-zero $\Delta \kappa$ in only the ordered phase and its strong field dependence.
\label{dkappa}}
\end{figure}

To shed light on the origin of the double-maximum structure of $\kappa(T)$, it is instructive to first look at $\kappa$ of the nonmagnetic homologue, LaRu$_2$Al$_{10}$.  The phononic contribution $\kappa_{\rm ph}(T)$ to the latter compound estimated by subtracting $\kappa_{\rm e}(T)$, as derived from $\rho(T)$ with the aid of the WF law, from the raw data, is shown in Fig.~\ref{kappa}. A comparison between the measured total thermal conductivity for CeRu$_2$Al$_{10}$ and $\kappa_{\rm ph}(T)$ of LaRu$_2$Al$_{10}$ reveals a good coincidence of their maxima at high temperature (40 $-$ 50 K), characteristic of lattice heat conductance in a crystalline solid. Furthermore, $\kappa(T)$ of single crystalline, isoelectronic CeFe$_2$Al$_{10}$ (dotted curve, adopted from ref. \cite{muro13} for $a$ axis) is also shown for comparison. This compound is characterized by a stronger hybridization between conduction electrons and $f$ states and, consequently, a nonmagnetic semiconducting ground state. Correspondingly, its $\kappa(T)$, which is dominated by the phononic contribution, reveals a single, phonon-derived maximum with smaller but comparable values. These facts suggest that the maximum at $T$ = 50 K in $\kappa(T)$ of CeRu$_2$Al$_{10}$ is phononic in origin, whereas the enhancement below $T_0$ is due to a different mechanism. Assuming a similar temperature dependence of $\kappa_{\rm ph}(T)$ for CeRu$_2$Al$_{10}$ to that of LaRu$_2$Al$_{10}$, one can straightforwardly extrapolate the curve of $\kappa(T)$ from above $T_0$ down to 2 K (dashed line in Fig.~\ref{kappa}). The excess thermal conductivity below $T_0$ for CeRu$_2$Al$_{10}$, denoted as $\kappa_{\rm m}$, is therefore obtained by subtracting the estimated $\kappa_{\rm ph}$ from the measured values of $\kappa$. As shown in Fig.~\ref{dkappa}a, in the ordered phase, $\kappa_{\rm m}$ amounts to sizable values that are comparable to $\kappa_{\rm ph}$.

Further evidence in support of an exotic mechanism rather than lattice vibration as origin of $\kappa_{\rm m}$ is found by applying magnetic fields:  While at $T$$>$$T_0$, $\kappa(T)$ is field insensitive, with the onset of the ordered phase at $T_0$, $\kappa_{\rm m}$ can be suppressed substantially by applying a magnetic field.  The suppression, revealed by the difference between $\kappa$ measured in zero and applied field, $\Delta \kappa$ $=$ $\kappa_{\rm 0T}$ $-$ $\kappa_{\rm B}$, is also shown in Fig.~\ref{dkappa}. The sudden development of $\Delta \kappa$ right below $T_0$ and its strong field dependence point to i) an intimate relationship of $\kappa_{\rm m}$ and the phase transition and ii) heat carriers responsible for $\kappa_{\rm m}$ are magnetic in origin or strongly coupled to magnetic excitations. Apart from phononic contribution discussed above, here one may also rule out structural change/distortion as a leading origin for $\kappa_{\rm m}$, given the field sensitivity of $\kappa_{\rm m}$.

As already mentioned, our observations on $\kappa(T)$ of CeRu$_2$Al$_{10}$ show striking resemblance to those of URu$_2$Si$_2$: In the hidden-order phase of the latter compound, a large enhancement of $\kappa$ as well as a field-induced suppression have been observed \cite{behnia05,sharma}. At $T$$>$$T_{\rm h}$, $\kappa(T)$ is insensitive to field, similar to our observations made for CeRu$_2$Al$_{10}$. The excess $\kappa$ below $T_{\rm h}$ has been interpreted as a consequence of sudden freezing out of scattering centers mainly for heat-carrying acoustic phonons \cite{sharma}. This conclusion is still under debate. Recent inelastic x-ray scattering measurements of phonon dynamics in URu$_2$Si$_2$ reveal that the phonon modes do not change significantly upon cooling into the hidden-order phase. This leads to the conclusion that $\kappa_{\rm ph}$ is much less important compared to that of magnetic excitations in the low temperature phase, despite the extremely small magnetic moment \cite{gardner}. Actually, if the excess thermal conductivity in the ordered phase would reflect a recovery of $\kappa_{\rm ph}(T)$ due to freezing out of some spin-lattice scattering events, one should expect $\kappa_{\rm ph}(T)$ to further increase upon applying magnetic field.  This is in contrast to the observations for CeRu$_2$Al$_{10}$ and URu$_2$Si$_2$, but is indeed the case for many multiferroic materials due to strong spin-lattice coupling \cite{xsun}. There, the enhancement of $\kappa_{\rm ph}(T)$ by applying a field results from suppression of magnetic fluctuations \cite{note}.

Results of inelastic x-ray scattering measurements suggest low-energy magnetic excitations to be the origin of the excess $\kappa$ in the hidden order phase of URu$_2$Si$_2$ \cite{gardner}. In spite of very tiny antiferromagnetic moments ($\sim$ 0.02 $\mu_{\rm B}$) detected in the hidden-order phase, well-defined magnetic excitations throughout the Brillouin zone have been indeed revealed by neutron scattering experiments for URu$_2$Si$_2$ \cite{neutron}. Likewise, our analysis of the thermal conductivity data support the same conclusion for CeRu$_2$Al$_{10}$, where heat transport by antiferromagnetic spin waves could to be the likely source of $\kappa_{\rm m}(T)$ below $T_0$. For an insulating spin system, this scenario has been well established and additional heat transport in its magnetically ordered phase has been proved to be an effective probe for low energy magnetic excitations \cite{insulator}. Along the same line, the significant $\kappa_{\rm m}$ in CeRu$_2$Al$_{10}$ and URu$_2$Si$_2$ is believed to benefit from the low carrier density of the ordered phase in these compound. However, it remains to be unraveled why a magnetic contribution to $\kappa(T)$ in a system with very small magnetic moments can be substantially large. In view of the high similarity of the thermal conductivity and close magnetic ordering temperature of CeRu$_2$Al$_{10}$ and CeOs$_2$Al$_{10}$ \cite{lue12}, it seems reasonable to believe that the above discussion applies to the latter compound as well.

\begin{figure}
\includegraphics[width=0.85\linewidth]{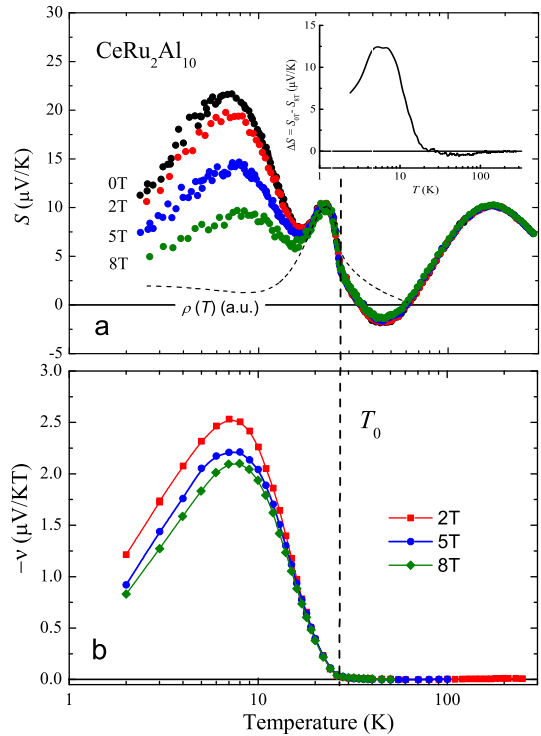}
\caption{(a) Thermopower $S(T)$ measured in different magnetic fields for CeRu$_2$Al$_{10}$. The curve of $\rho(T)$ is also shown in order to highlight the similarity of the sharp peak at 22 K in both $S(T)$ and $\rho(T)$. Inset: The magneto-thermopower defined $\Delta S$ = $S_{\rm 8T}$ $-$ $S_{\rm 0T}$. (b) Nernst coefficient $\nu$ measured in different fields shown as $-\nu$ vs $T$.
\label{tep}}
\end{figure}

In Fig.~\ref{tep}a,b we show the thermopower $S(T)$ and Nernst coefficient $\nu(T)$ measured in different magnetic fields for CeRu$_2$Al$_{10}$. $S(T)$ measured at $B$ = 0 T is characterized by two broad maxima at $T$ = 7 K and 180 K, together with a sharp peak emerging in between, at $T$ $\approx$ 22 K, in agreement with previous reports \cite{strydom09, muro10}. On the other hand, $|\nu(T)|$ is dramatically enhanced by two orders of magnitude with the onset of the ordered phase below $T_0$. Except for the sharp peak at 22 K, a double-maximum feature of $S(T)$ had been frequently observed for Kondo systems, e.g.,  CeRu$_2$Si$_2$ \cite{amato}, reflecting the Kondo scattering on  the crystalline-electric-field (CEF) derived Kramers ground-state and excited doublets. Indeed, the first excited doublet of CeRu$_2$Al$_{10}$ in the orthorhombic CEF is about 30 meV above the ground state \cite{cef}, in accordance with the position of the high temperature maximum of $S(T)$. Interestingly, the values of $S(T)$ remain the same magnitude while the carrier concentration changes by two orders magnitude for the two phases separated by $T_0$ \cite{strydom09}. This is the case for URu$_2$Si$_2$ \cite{bel}, too. For Kondo systems, such observation is not surprising because $S$ is strongly dependent on the Kondo scattering rather than the electronic density of states (DOSs) of the conduction band \cite{sun13}. Upon applying magnetic fields, at $T$ $<$ 20 K the values of $S(T)$ commence to become reduced, whereas $S(T)$ remains unchanged at higher temperatures, as revealed by $\Delta S$ (= $S_{\rm 0T}$$-$$S_{\rm 8T}$) as a function of temperature (Fig. 4a, inset). These distinctly different field responses of thermopower point to two different Kondo energy scales below and above $T_0$, related to the abrupt change of the quasi-particle DOSs by opening of charge gap. $S(T)$ varies with temperature sublinearly below the low-temperature maximum, with an initial slope $S/T$ = 4.8 $\mu$V/K$^2$ at $B$ = 0 T. Given the electronic specific-heat coefficient $\gamma$ = 24.5 mJ/mol K$^2$ in the ordered phase \cite{nishioka09}, the dimensionless ratio of $S/T$ and $\gamma$, i.e., $q$ = $S$/$\gamma$$T$, amounts to 18.8 for CeRu$_2$Al$_{10}$. This value, as well as $q$ = 4.5 for URu$_2$Si$_2$ \cite{behnia04}, is larger than $q$ ($\approx$ 1) of most heavy-fermion metals, consistent with their low-carrier nature in the ordered phase.

The sharp $S(T)$ peak at $T$ $\approx$ 22 K appears to signify the temperature regime where both $\mu_H(T)$ and $\nu(T)$ (Fig. 4b) show pronounced changes. Due to its field insensitivity this peak cannot be ascribed to a magnon-drag effect associated with electron-spin wave interaction. We recently have demonstrated for the electron-doped skutterudite CoSb$_3$ that an anomalous thermopower can stem from a dramatic change of charge mobility with respect to temperature \cite{cosb3}. While the thermopower is commonly referred to energy dependence of the electronic DOSs at the Fermi level, it may well derive from a significantly energy-dependent relaxation time $\tau$ as evidenced by a largely temperature-dependent $\mu_{\rm H}$.  Following this description, the $S(T)$ peak at 22 K can be qualitatively described by the derivative of Hall mobility, $d\mu_H/dT$. For details, see supplementary information of ref. \cite{cosb3}. The fact that this anomalous $S(T)$ peak sits right on top of the $\rho(T)$ maximum (cf. Fig. 4a) lends further support to this argument. Noticeably, the absolute value $|S(T)|$ of URu$_2$Si$_2$ increases abruptly below $T_{\rm h}$ as well \cite{bel}, which may be partially derived from the strong variation of $\mu_{\rm H}(T)$ as discussed above.  When magnetic field is applied, even a two-maximum structure in $S(T)$ is observed below $T_{\rm h}$ \cite{bel}, mimic to the case of CeRu$_2$Al$_{10}$.

The Nernst coefficient $|\nu(T)|$, which increases by more than two orders of magnitude with the onset of the low-$T$ phase in both the current system (Fig. 4b) and URu$_2$Si$_2$ \cite{bel,naturep} deserves special attention as well. Within the Boltzmann theory, one indeed expects a large Nernst signal from high mobility charge carriers: In a first-order approximation, $\nu$ $\propto$ $\frac{k_{\rm B}T}{\epsilon_{\rm F}}$$\mu_{\rm H}$ \cite{bel}, with $k_{\rm B}$ being Boltzmann constant and $\epsilon_{\rm F}$ the Fermi energy. Nevertheless, the microscopic mechanism for large Nernst signal in these compounds may be not that straightforward. For URu$_2$Si$_2$, additionally, the large Nernst response has been discussed as a fingerprint of possible chiral or Berry-phase fluctuations associated with the broken time-reversal symmetry of the superconducting order parameter \cite{naturep}. Finally, it is instructive to mention that, apart from transport signatures, analogies between the two systems have also been found by other probes, such as optical conductivity, which exhibits a gap structure with a mysterious charge excitation peak \cite{kimura}, and NQR/NMR spectra, which reveal a similar nuclear-spin lattice relaxation rate for both compounds \cite{matsu09}.

To summarize, we have investigated various magneto-transport properties of the Kondo semiconducting compound CeRu$_2$Al$_{10}$ and stressed their analogies with those of the hidden-order compound URu$_2$Si$_2$. With the onset of the ordered phase at $T_0$ = 27 K, CeRu$_2$Al$_{10}$ experiences a dramatic change of charge dynamics exhibiting very different charge mobilities below and above $T_0$. This has profound effects on the electrical and thermoelectrical responses, leading to an additional peak in both $\rho(T)$ and $S(T)$, as well as an enhanced Nernst coefficient $\nu(T)$, resembling the case of URu$_2$Si$_2$. Most significantly, in the ordered phase, thermal conductivity exhibits an excess contribution presumably derived from low-energy magnetic excitations and sensitive to magnetic field, in strong parallel to URu$_2$Si$_2$ as well. Given the Kondo semiconducting behavior above $T_0$, which usually involves a nonmagnetic ground state, it is surprising that this compound shows magnetic order with small moments at a surprisingly high temperature. Like the hidden order phase in URu$_2$Si$_2$, where the tiny antiferromagnetic moments can not be the main order parameter, the low-temperature magnetic phase in CeRu$_2$Al$_{10}$ remains to be elusive and badly calls for further work.

This work was supported by the MOST of China (Grant No: 2015CB921303, 2015CB921304), the National Science Foundation of China (Grant No:11474332), and the Chinese Academy of Sciences through the strategic priority research program (XDB07020200). A.M.S. thanks the SA-NRF (93549) and the URC/FRC of UJ for financial assistance.


\begin{thebibliography}{99}

\bibitem{strydom09} A.M. Strydom, Phys. B {\bf 404}, 2981 (2009).
\bibitem{nishioka09} T. Nishioka, Y. Kawamura, T. Takesaka, R. Kobayashi, H. Kato, M. Matsumura, K. Kodama, K. Matsubayashi, and Y. Uwatoko,
 J. Phys. Soc. Jpn. {\bf 78}, 123705 (2009).
\bibitem{Kha10} D.D. Khalyavin, A.D. Hillier, D.T. Adroja, A.M. Strydom, P. Manuel, L.C. Chapon, P. Peratheepan, K. Knight,
P. Deen, C. Ritter, Y. Muro, and T. Takabatake, Phys. Rev. B {\bf 82}, 100405(R) (2010).
\bibitem{kato} H. Kato, R. Kobayashi, T. Takesaka, T. Nishioka, M. Matsumura, K. Kaneko and N. Metoki, J. Phys. Soc. Jpn. {\bf 80}, 073701 (2011).
\bibitem{behnia05} K. Behnia, R. Bel, Y. Kasahara, Y. Nakajima, H. Jin, H. Aubin, K. Izawa, Y. Matsuda, J. Flouquet, Y. Haga, Y. \=Onuki and P. Lajey, Phys. Rev. Lett. {\bf 94}, 156405 (2005).
\bibitem{sharma} P.A. Sharma, N. Harrison, M. Jaime, Y.S. Oh,  K.H. Kim, C.D. Batista, H. Amitsuka and  J.A. Mydosh, Phys. Rev. Lett. {\bf 97}, 156401 (2006).
\bibitem{gardner} D. R. Gardner, C. J. Bonnoit, R. Chisnell, A. H. Said,  B. M. Leu,  T. J. Williams, G. M. Luke and  Y. S. Lee, Phys. Rev. B {\bf 93}, 075123 (2016).
\bibitem{cosb3} P. Sun, B.P. Wei, J.H. Zhang, J.M. Tomczak, A.M. Strydom,  M. S\o ndergaard, B.B. Iversen, and F. Steglich, Nat. Commn. {\bf 6}, 7475 (2015).
\bibitem{matsu09} M. Matsumura, Y. Kawamura, S. Edamoto, T. Takesaka,  H. Kato,  T. Nishioka, Y. Tokunaga, S. Kambe, and  H. Yasuoka,
J. Phys. Soc. Jpn. {\bf 78}, 123713 (2009).
\bibitem{kimura} S. Kimura, H. Tanida,  M. Sera,  Y. Muro,  T. Takabatake, T. Nishioka, M. Matsumura, and  R. Kobayashi, Phys. Rev. B {\bf 91}, 241120(R) (2015).
\bibitem{dawson} A. LeR Dawson, W.R. Datars,  J.D. Garrett, and F.S. Razavi, J. Phys.: Condens. Matter {\bf 1}, 6817 (1989).
\bibitem{tanida10} H. Tanida, D. Tanaka, M. Sera, C. Moriyoshi, Y. Kuroiwa, T. Takesaka, T. Nishioka, H. Kato, and M. Matsumura, J. Phys. Soc. Jpn. {\bf 79}, 063709 (2010).
\bibitem{lue12} C.S. Lue,  H.F. Liu, B.D. Ingale, J.N. Li, and Y. K. Kuo, Phys. Rev. B {\bf 85}, 245116 (2012).
%\bibitem{lue10} C.S. Lue, S.H. Yang, A.C. Abhyankar, Y.D. Hsu, H.T. Hong, and Y.K. Kuo, Phys. Rev. B {\bf 82}, 045111 (2010).
\bibitem{muro13} Y. Muro, K. Yutani, J. Kajino, T. Onimaru, and T. Takabatake, J. Korean Phys. Soc.  {\bf 63}, 508 (2013).
\bibitem{xsun} X. M. Wang, C. Fan,  Z. Y. Zhao, W. Tao, X.G. Liu, W.P. Ke, X. Zhao, and X. F. Sun, Phys. Rev. B {\bf 82}, 094405 (2010).
\bibitem{note} In the vicinity of a critical field where a magnetic phase transition takes place, thermal conductivity can be suppressed by field-induced critical fluctuations, cf. ref. \cite{xsun}.
\bibitem{neutron} F. Bourdarot, S. Raymond, and L.-P. Regnault, Philos. Mag. {\bf 94}, 3702 (2014).
\bibitem{insulator} See, for example, R. Jin $et$ $al.$, Phys. Rev. Lett. {\bf 91}, 146601 (2003).
\bibitem{muro10} Y. Muro, K. Motoya, Y. Saiga, and T. Takabatake, J. Phys.: Confer. Series {\bf 200}, 012136 (2010).
\bibitem{amato} A. Amoto, D. Jaccard, J. Sierro, F. Lapierre, P. Haen, P. Lejay, J. Flouquet, J. Magn. Magn. Mater. {\bf 76\&77}, 263 (1988).
\bibitem{cef} F. Strigari $et$ $al.$, Phys. Rev. B {\bf 86}, 081105(R) (2012).
\bibitem{bel} R. Bel, H. Jin, K. Behnia, J. Flouquet, and P. Lejay, Phys. Rev. B {\bf 70}, 220501(R) (2004).
\bibitem{sun13} P. Sun and F. Steglich, Phys. Rev. Lett. {\bf 110}, 216408 (2013).
\bibitem{behnia04} K. Behnia, D. Jaccard, and J. Flouquet,  J. Phys.: Condens. Matter {\bf 16}, 5187 (2004).
\bibitem{naturep} T. Yamashita, Y. Shimoyama, Y.Haga, T.D. Matsuda, E. Yamamoto,Y.\=Onuki, H. Sumiyoshi, S. Fujimoto, A. Levchenko, T.Shibauchi, and Y. Matsuda, Nat. Phys. {\bf 11}, 17 (2015).


\end{thebibliography}
\end{document}